\documentclass[aps,a4paper,twocolumn,showpacs]{revtex4-1}
\usepackage{graphicx}
\usepackage{latexsym}
\usepackage{amsbsy}

\newcommand{\kB}{k_{\mathrm{B}}}
\newcommand{\om}{{ \mbox {\boldmath  $\omega$} }}

\newcommand{\fettmu}{{ \mbox {\boldmath  $\mu$} }}

\usepackage{epstopdf}

\newcommand{\fett}[1]{\mbox{\boldmath$#1$}}

\begin{document}

\date{\today}

\title{Dynamics of magnetic nano particles  in a viscous fluid driven by rotating magnetic fields} 

\author{Klaus \ D.\ Usadel}

\affiliation{  Theoretische Physik,
  Universit\"{a}t Duisburg-Essen, 47048 Duisburg, Germany}

\begin{abstract}
The rotational dynamics of  magnetic nano particles in rotating magnetic fields  in the presence of thermal noise is studied both theoretically and by performing numerical calculations. 
 Kinetic equations for the dynamics of particles with uniaxial magnetic anisotropy are studied and the phase lag between the rotating magnetic moment and the driving field is obtained. It is shown that  for large enough anisotropy energy  the magnetic moment is locked to the anisotropy axis so that the particle behaves like a rotating magnetic dipole. The corresponding rigid dipole model is analyzed both numerically by solving the appropriate Fokker-Planck equation  and analytically by applying an effective field method. In the special case of a rotating magnetic field  applied analytic results are obtained in perfect agreement with numerical results based on the Fokker-Planck equation.  The analytic formulas derived are not restricted to small magnetic fields   or low frequencies and are therefore important for applications.   
The illustrative numerical calculations presented are performed for magnetic parameters typical for iron oxide.

  \end{abstract}
\pacs{75.40.Gb, 75.40.Mg, 75.75.Jn, 75.60.Jk } \maketitle

\section{Introduction} 
\label{s:intro}

 Recently, magnetic nanoparticles (MNPs) have been studied for many biomedical applications such as magnetic particle imaging,  separation of biological targets, immunoassays, drug delivery, and hyperthermia treatment \cite{pank,pank2,li}. In these applications  the magnetic moment may rotate within the particle with respect to some crystal axis - i.e. Ne\'{e}l relaxation - and move along with the particle with respect to the liquid, i.e. Brownian relaxation. Hyperthermia, for instance,  is based on the fact that power is absorbed locally by the MNP when placed in an applied oscillating magnetic field.  A  high absorption rate is achieved if both  Ne\'{e}l relaxation and Brownian relaxation contribute  \cite{usov_0,usa_ROT,kaf}. 

On the other hand if the anisotropy energy is large compared to the thermal energy  Ne\'{e}l relaxation becomes unimportant. In this limit  the MNP can be considered as a rigid body (RB) having a constant magnetic moment firmly attached to it. In this  rigid dipole model (RDM) the external driving field will create a torque to the magnetic moment which is transferred to the RB. In a rotating magnetic field, for instance,  the MNP will rotate as well.  
Theoretically the dynamics of MNPs has been studied in recent years in very many papers 
for the case that the particles are fixed in space.  
The dynamics is then reduced to that of  magnetic moments in external fields for which the
stochastic Landau-Lifshitz-Gilbert (LLG) equation as introduced by Brown
\cite{brown} is often used.  This approach has gained increased interest recently  because
of its application to magnetic single-domain particles. In these MNPs the magnetic moments within the particles are firmly tied together making it possible to describe the magnetic moment of the  particle as one macro spin of
constant length \cite{Ohand00}. Its dynamics is expected to be well described by a classical
approach. 
The dynamics of these macro spins is greatly influenced by thermal fluctuations originated from coupling of the spins to the surrounding medium (the heat bath). 
Additionally, for mobile particles  Brownian relaxation has to be taken into account \cite{ shliomis_0, coffey, shliomis, lahonian, nedelcu, rosen, usov, mam, usa_ROT}. 

Recently the dynamical properties of mobile particles have been studied in some detail in connection with biological applications, i. e.  the realization of homogeneous biosensors \cite{conn, chung, enpuku, eber} based on  the response of a suspension of MNPs to an applied ${\it ac }$ magnetic field.  In these  biological applications a dilute suspension of MNPs is present  the physical properties of which are determined to a large extend by the magnetic and mechanical dynamics of independent MNPs. 

Quite recently  biosensors were proposed \cite{schritt, schritt2, dieck} which are based on the response of MNPs with Brownian dynamics to rotating magnetic fields, i.e.  to differences  in  the phase lag between  the rotating magnetic moment and the  driving field in order to determine the changes in hydrodynamic volume caused by analytes  bound to the surface of the MNPs. 
 For this method to be successful, it is necessary to quantitatively clarify the dynamics of MNPs in rotating fields. Theoretical studies on this problem reported in Ref. \cite{ferro} are based on an effective field method (EFM) \cite{mart, raik} for the dynamics of an ensemble of MNPs treated as  rigid dipoles placed in a viscous medium. The EFM was derived from a Fokker-Planck equation (FPE)  describing the rotational dynamics at finite temperatures.

Results obtained for the case that the applied field rotates were compared with measurements of the  phase lag \cite{dieck, dieck2} 
revealing  a reasonable agreement between the measured and calculated values in the low frequency regime and for small field amplitudes. However, significant discrepancies were observed outside this regime. Numerical calculations  based directly on the FPE  for the rigid dipole model \cite{yosh} confirmed these findings  so that a need for further theoretical work persists. 

In the present paper we therefore study the dynamics of mobile MNPs driven  by a rotating magnetic field in order to contribute to an understanding of the physics underlying the functionality of the class of biosensors mentioned \cite{schritt,schritt2, dieck}.  
The ensemble of MNPs is described by a set of kinetic equations proposed earlier \cite{usa_ROT} for the dynamics of mobile particles treating Brownian and Ne\'{e}l dynamics on equal footing. 

In the first part of the paper we discuss briefly our kinetic equation  \cite{usa_ROT} and present numerical results for the dynamics of MNPs with finite uniaxial anisotropy driven by rotating magnetic fields. Evidence is given to the expected blocking of the magnetic moment to the anisotropy axis for large enough anisotropy energies. The relation between our kinetic equations and the RDM is discussed and it is shown how the RDM emerges from our kinetic equations. 

In the second part of the paper we discuss in detail the RDM and sketch the derivation of the  basic equations of the EFM  \cite{mart, raik}. 
We show that for the special case of an applied rotating magnetic field these equations can be solved without further assumptions.  Analytic formulas are obtained for the phase lag and for the magnetic moment and it is shown that these results are in nearly perfect agreement with results which follow from numerical solutions of the FPE.  The analytic formulas derived are not restricted to small magnetic fields and/or frequencies  and are therefore important for applications.

\section{Dynamics of magnetic nano particles with finite anisotropy}

\subsection{Kinetic equations}

A nano particle  considered in the present paper is modeled as a uniformly magnetized spherical rigid body (RB) with uniaxial magnetic anisotropy. The particle can rotate in a viscous medium. Its orientation  in space is described by a time dependent unit vector ${\bf n}(t)$ parallel to the anisotropy axis of the RB. This vector is firmly attached to the RB so that its equation of motion is given by 
 \begin{equation}
\frac{d {\bf n}}{ dt} =  \om \times \bf n 
\label{e:n_punkt}
\end{equation}
where $\om$ denotes the  angular velocity of the RB. 
The differentiation in Eq. (\ref{e:n_punkt}) is performed in a coordinate system fixed in space, the laboratory frame.

The magnetic moment $\fettmu$ of the NP with constant magnitude  $\mu_s$ can rotate in space along  with the RB and can also rotate relative to it depending primarily on the strength of the anisotropy energy $D$ of the NP.  We introduce  $\bf e$, an unit vector in the direction of the magnetic moment, $ \bf e = \fettmu / \mu_s$. For $\bf e$ and for the angular velocity $\om$ of the RB   kinetic equations were proposed in  \cite{usa_ROT} which  can be written in compact form as 
\begin{eqnarray}
\frac{d {\bf e}}{d t} = \om \times \bf e - \frac {\gamma}{1+ {\alpha}^2  } \Big (\bf e \times {(\mathbf B}_{e} -\frac{1}{\gamma} \om) \nonumber \\ +  \alpha \,\bf e\times (\bf e \times { (\mathbf B }_{e} -\frac{1}{\gamma}  \om))  \Big)
\end{eqnarray}
and  
\begin{eqnarray}
 \Theta \, \frac{d {\om}}{d t}= \frac {\mu_s}{\gamma} \frac{d \bf e}{d t} + \mu_s \bf e \times (\bf B + \boldsymbol \zeta) - \xi \, \om +\fett \epsilon .
\label{e:omega_punkt_2}
\end{eqnarray}
The effective field ${\bf B}_e$ entering Eq.(2) consists of the external driving field ${\bf B}$  which may be time dependent,  a term proportional to the anisotropy energy $D$ and a stochastic field $\boldsymbol \zeta$  relevant at elevated temperatures, 
\begin{equation}
{\bf B}_{e}= {\bf B}+  \frac {2 \,D}{{\mu}_s} \, (\bf e \cdot \bf n )  \,  \bf n + \boldsymbol \zeta . 
\end{equation}

The quantity $\alpha$ appearing in Eq.(2) is the dimensionless damping parameter usually used in the literature \cite{garcia, nowak, usad_1}, $\xi$ denotes  the friction coefficient  usually expressed as $\xi = 6 \eta V_{d}$ where  $\eta$ is the dynamical viscosity while $V_{d}$ with radius $r_d$ denotes the hydrodynamic or total volume of the particle, i.e. it is assumed that the particle  consists of a magnetic core region of volume $V_m$ with radius $r_m$ eventually covered by a nonmagnetic surfactant layer.

For the thermal fluctuations introduced above it is assumed as usual that they are Gaussian distributed with zero mean. Their 
correlators  are chosen in such a way that in equilibrium Boltzmann statistics is recovered. This leads to\begin{equation}
  \langle \zeta_{l}(0)\zeta_{m}(t) \rangle_\zeta =
   \delta_{l,m} \delta(t) 2 \alpha k_B T / (\mu_s \gamma).
  \label{e:corr4}
\end{equation}
for the magnetic field fluctuations \cite{brown} and
\begin{equation}
  \langle \epsilon_{l}(0)\epsilon_{m}(t) \rangle_\epsilon =  \delta_{l,m} \delta(t) 2 \, \xi \, k_B \,T .
  \label{e:corr5}
\end{equation}
for the  Brownian rotation \cite{coffey, coffey_kal}.  The angular brackets denote averages over $\fett \zeta$ and $\fett \epsilon$, respectively, and  $l$ and $m$ label cartesian components of the fluctuating fields. 

For given fluctuating fields ${\fett \epsilon}(t)$ and ${\fett \zeta}(t)$ the quantities ${\bf n}(t)$ and ${\bf e}(t)$ as solutions of the stochastic equations are trajectories on the unit sphere because the equations of motion conserve the length of these vectors. Physical quantities of interest describing properties of an ensemble of identical particles are obtained as averages over these trajectories. The reduced magnetic moment at finite temperatures, for instance, is given by 
\begin{equation}
{\bf m}(t) = \langle{\bf e}(t)\rangle_{\zeta, \epsilon}.
\end{equation}

Eqs. (1 - 6 ) constitute a closed set of kinetic equations for the quantities ${\bf n}(t)$, ${\bf e}(t)$ and ${ \om}(t)$  specifying the dynamics of the MNP.  In deriving these equations we were guided by the requirement that for the isolated particle the total angular momentum $ \bf L + \bf S $ has to be conserved irrespectively of internal interactions within the MNP. Here,  ${\bf L} = \Theta  \om$   denotes the angular momentum and $\bf S$ the spin momentum, $\bf S =  - {\gamma}^{-1} \, \fettmu $, where  $\gamma$ denotes  the gyromagnetic ratio. For more details we refer to \cite{usa_ROT}.

The kinetic equations proposed treat the dynamics of the magnetic moment and the rotational motion of the NP on the same footing. In general, both processes are coupled  leading to a rather complex behavior. In the limit of a large anisotropy energy, however, it is expected that the magnetic moment is locked into a position parallel to the anisotropy axis of the RB.  This limiting case is the essence of the rigid dipole model. 
\subsection{Numerical analysis: dependence on anisotropy energy}

Using our kinetic equations we investigate numerically the dependence of phase lag and induced magnetic moment on the anisotropy energy considering the anisotropy constant $K_1$ as being adjustable. Other parameters are chosen as  typical for particles of iron oxides (magnetite).  The magnetic  moment ${\mu}_s$ is expressed as  ${\mu}_s = M_s \, V_m$ with $M_s=4 \cdot 10^5 A/m$ and the reduced viscosity $\tilde \eta$ is defined as  $\tilde \eta = \eta / \eta_{water} $ with $\eta_{water} = 10^{-3} \, $ kg/ms.
Results are expressed as function of $K_1/K_{10}$ with  $K_{10} = 10^4\,J/m^3$ which orresponds to the  anisotropy constant of magnetite.
 The rotating magnetic field is given by
\begin{equation}
 {\bf B}  = B_0 \,({\mathrm {cos}}( \tilde \omega \,t) \,{\hat {\bf x}} + {\mathrm {sin}}( \tilde \omega \,t) \,{\hat {\bf y}} )
 \label{e:t}
 \end{equation}
 where $\tilde \omega = 2\pi F$ denotes the frequency of the driving field. A cartesian coordinate system fixed in space is used  spanned by unit vectors  ${\hat {\bf x}}$, ${\hat {\bf y}}$ and ${\hat {\bf z}}$.
 
\begin{figure}[h] 
   \centering
 \includegraphics[width=3.0in]{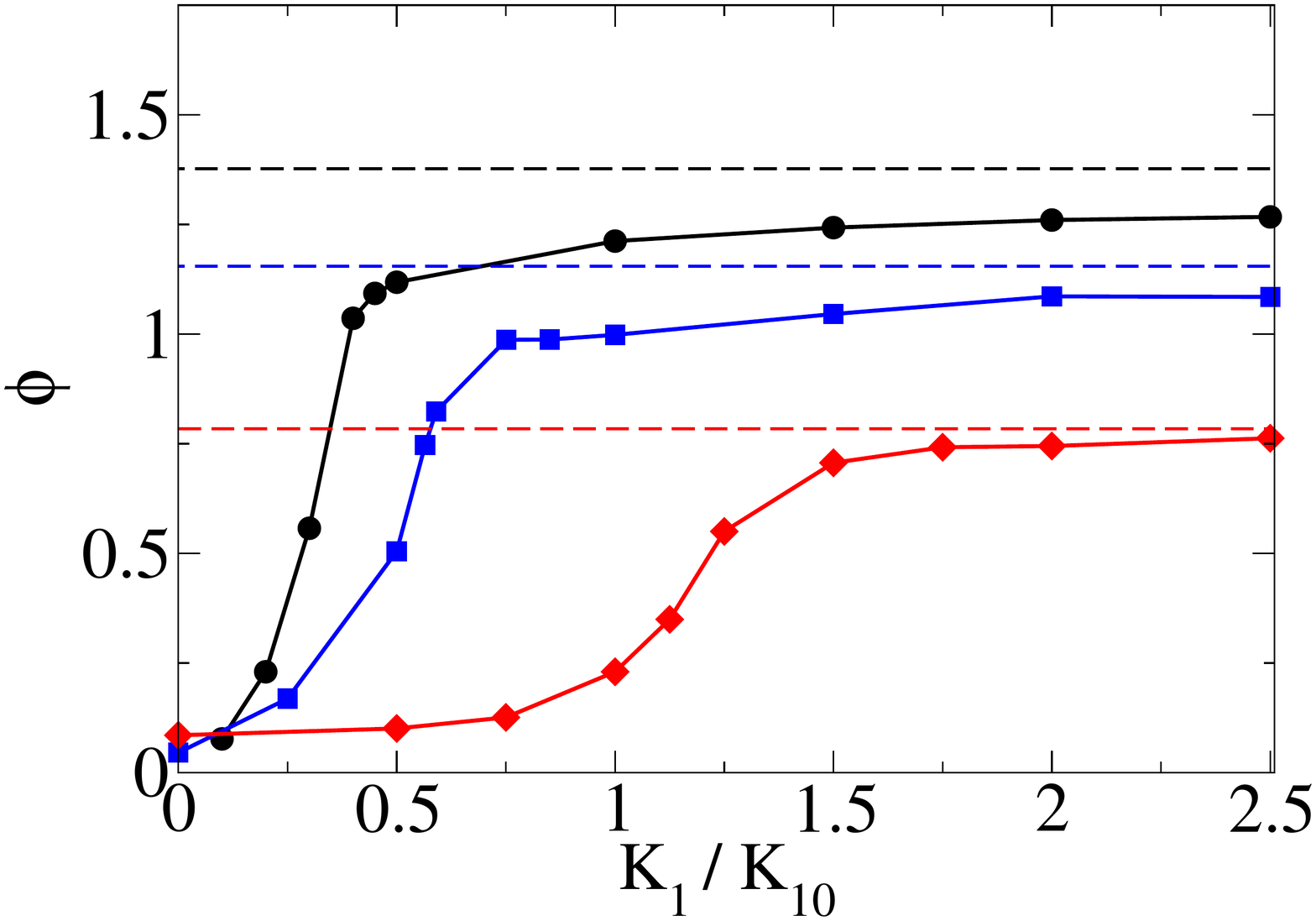}  \\[-0.3cm]
  \includegraphics[width=3.0in]{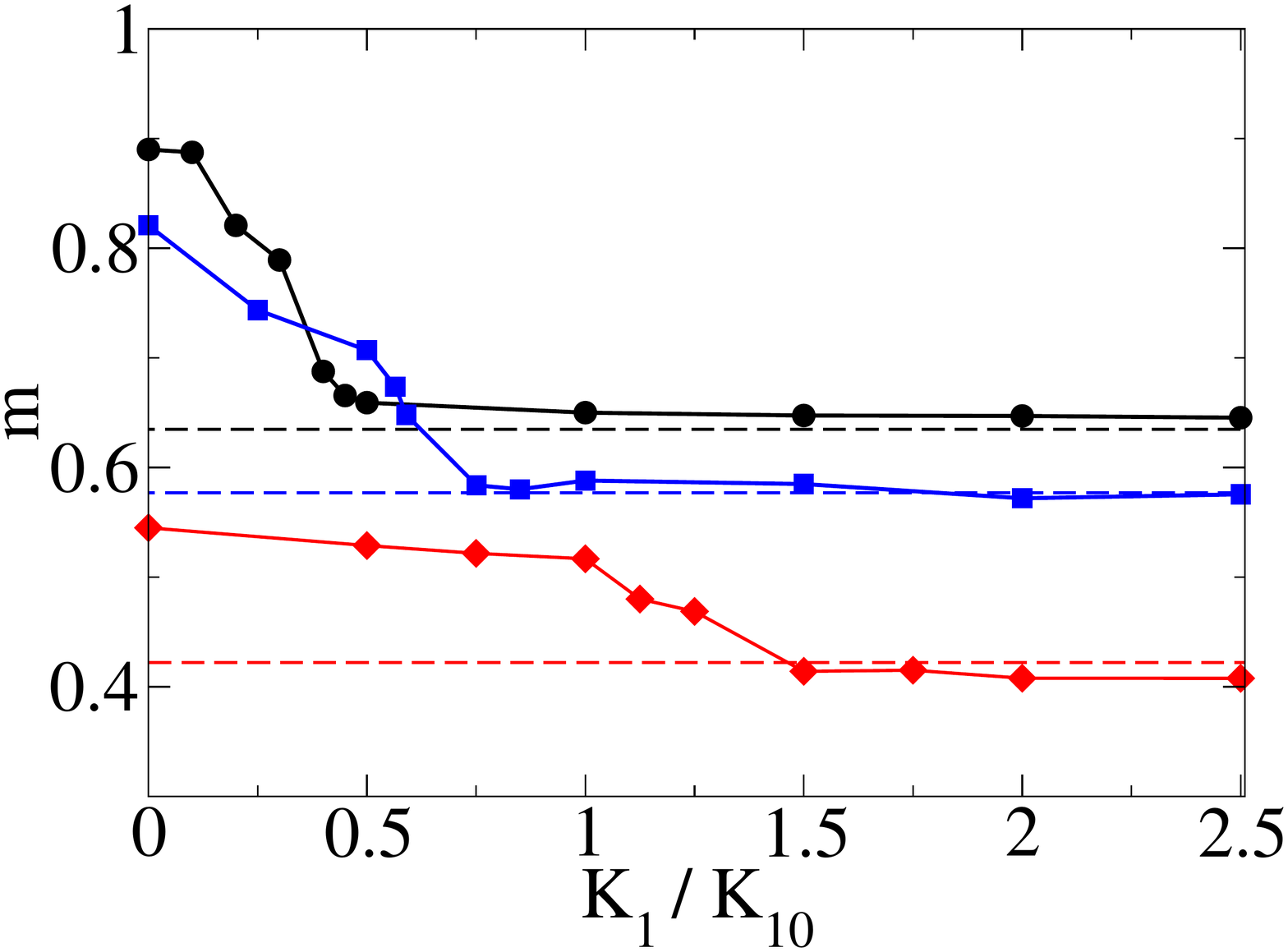}  \\[-0.3cm]
\caption{(color online) Phase lag  $\phi$ ( upper panel ) and in-plane magnetization $m$ ( lower panel ) versus reduced anisotropy parameter $K_1/K_{10}$ for different values of the magnetic radius $r_m$: $r_m= 19$ nm (circles, black), $14$ nm (squares, blue), and $10$ nm (diamonds, blue). 
 $\alpha = 0.01$, $\tilde \eta = 1.0$, $r_d/r_m   = 2.5$, $F=4$ kHz, $B=5$ mT. The horizontal lines indicate the values obtained within the RDM.}      
   \label{f:fig_1}
   \end{figure}  

The initial conditions necessary for the numerical integration of the kinetic equations are specified as  
${\bf e}(t=0) = {\bf n}(t=0)$
and $\om (t=0)=0$ with randomly distributed    ${\bf e}(t=0)$. For the integration of the stochastic equations Eqs.(1-3) methods well known by now from literature are used \cite{garcia, usad_1,nowak}. In the stationary state which is reached after a  time interval $\delta t$  the ensemble averaged quantities $\langle{\bf e}(t)\rangle_{\zeta,\epsilon}$  and $\langle{\bf n}(t)\rangle_{\zeta,\epsilon}$ are monitored.  Note that $\delta t$ depends very much on parameters.  
 
In the stationary state $\langle {\bf e}(t) \rangle_{\zeta,\epsilon}$  rotates around  $\hat {\bf z}$ with the frequency of the driving field.  A phase lag is observed between the direction of the rotating field and this rotating magnetic moment. The   phase lag $\phi$  is  obtained as the angel between  $\langle{\bf e}(t)\rangle_{\zeta,\epsilon}$ and the direction of the driving field  averaged over about $1000$ particles in the ensemble followed by an average over about $100$ field cycles.              
   Fig. 1 shows the averaged phase lag  as function of $K_1/K_{10}$ for three different magnetic radii $r_m$ of the MNP, $r_{m}$ $= 10$ nm (circles, black),$14$ nm (diamonds, blue) and $19$ nm (squares, red). 
   
   For large anisotropy energies it is observed that the phase lag becomes independent of the anisotropy energy which indicates that the moment is locked to the RB, i.e. it behaves like a rotating rigid dipole. This is  directly confirmed by the observation that   $\langle{\bf n}(t)\rangle_{\zeta,\epsilon}$ remains nearly parallel to $\langle {\bf e}(t) \rangle_{\zeta,\epsilon}$ (not shown), i.e. magnetic moment and anisotropy axis rotate unisono. 
 
The RDM is studied in detail in the next sections. Numerical results will be obtained some of which are shown as dashed horizontal lines in Fig.(1). They represent phase lag and magnetic moment, respectively, calculated within the RDM. Obviously for large  $K1/K10$ the values obtained from the RDM are approached asymptotically. 
 
       A reduction of $K_1/K_{10}$, on the other hand, leads to a sharp drop in the phase lag accompanied by a significant increase of $m$. In this limit the moment adjusts nearly parallel to the driving field and rotates with the field while the RBs stop rotating in a coherent fashion.  This follows from the time variation of  $\langle{\bf n}(t)\rangle_{\zeta,\epsilon}$ (not shown) consisting of only small random variations around zero.

In the course of this work we kept the temperature fixed at $T=300$ K. A variation of $T$ will effect in particular the Ne\'{e}l relaxation time which depends on $K_1/T$. We therefore expect that the value of  $K_1$ at which the locking of the moment sets in will depend on $T$.  
 
The increase of $m$ when reducing the anisotropy energy sets in for values of $K_1/K_{10}$ for which  the magnetic moment  unlocks from the RB moving towards the direction of the field. In the extreme limit $K_1=0$ the magnetic moment is only subjected to a tiny frictional torque resulting from the Gilbert - damping $\alpha$. Only for unrealistic high values of $\alpha$ this torque is large enough to overcome the viscous damping necessary for the RB to rotate.   
\subsection{Large anisotropy energy}

The numerical results presented suggest that for large anisotropy energies our kinetic equations pass over to the RDM.  This is supported  by the following  qualitative arguments. Consider a rotating frame  firmly attached to the RB in which the anisotropy axis is fixed. The time derivative $\frac {d \bf e}{dt}$ can be split into two parts, 
\begin{equation}
\frac {d \bf e}{dt} = \om \times {\bf e} + (\frac {d  \bf e}{d t})_{rf}
\end{equation}  
where the second term in this equation is the time derivative in the rotating frame. Comparison with Eq.(2) shows that this time derivative is equal to the second term of Eq.(2) so that
\begin{eqnarray}
(\frac{ {d \bf e}}{d t})_{rf} = - \frac {\gamma}{1+ {\alpha}^2  } \Big (\bf e \times {(\mathbf B}_{e} -\frac{1}{\gamma} \om) \nonumber \\ +  \alpha \,\bf e\times (\bf e \times { (\mathbf B }_{e} -\frac{1}{\gamma}  \om))  \Big), 
\end{eqnarray}
describing the dynamics of a spin with  a LLG-type equation with  effective field ${\bf B}_e - \frac{1}{\gamma}\om$ in the rotating frame. In this rotating frame the anisotropy axis is fixed. Therefore $\bf e$ will  settle nearly instantaneously parallel to the anisotropy axis and will stay there providing the anisotropy energy is much larger than the contributions from the driving field  and from $\om$ and that the temperature is such that thermal switching of $\bf e$ does not take place. This means that after a (short) initial time  ($\frac{d {\bf e}}{d t})_{rf}$  will go to zero and we are left with
\begin{equation}
\frac{d {\bf e}}{ dt} =  \om \times \bf e.
\label{e:e_punkt}
\end{equation}
Note that this argument requires a finite Gilbert damping $\alpha$. If the damping parameter is strictly zero there will be no relaxation towards the anisotropy axis so that  a rather complex precessional motion results.

Thus for large anisotropy energy we are lead to Eqs. (3) and (11) which constitute the basis of the RDM. Further simplifications are possible. First we note that in the equation of motion for $\om$ inertia effects can be ignored because of the small particle size combined with realistic values of $\xi$ so that one obtains
\begin{eqnarray}
\om = \frac {1}{\xi}  \, (\frac {\mu_s}{\gamma} \frac{d \bf e}{d t} + \mu_s \bf e \times (\bf B+\boldsymbol \zeta) + \boldsymbol \epsilon) 
\label{e:omega_punkt_2}.
\end{eqnarray}
Eqs.(11-12)  can be solved for $\frac{d \bf e}{dt}$ with algebraic manipulations known from the LLG-equation. The resulting equation for $\frac{d \bf e}{d t}$  depends on the parameter $\kappa$, 
\begin{eqnarray}
 \kappa = \frac{\mu_s}{\xi \gamma} 
\label{e:omega_punkt_2}
\end{eqnarray}
which is extremely small. It is suffice therefore to keep only the leading terms with respect to $\kappa$ resulting in the well-known  equation of motion for the dynamics of a magnetic NP in the rigid dipole approximation, 
\begin{equation}
\frac{d \bf e}{d t} = - {\xi^{-1}} \mu_s\,  \bf e \times (\bf e \times { B}) - {\xi ^{-1}}  \bf e \times { \boldsymbol \epsilon }.
\end{equation}
Note that the fluctuating field $\boldsymbol \zeta$ entering Eq.(12) can be shown to be negligible in lowest order in $\kappa$.

\section{Rigid dipole model}

The rigid dipole model became a matter of particular interest  quite recently because of its potential for biomedical applications. The stochastic equation describing its dynamics at finite temperature, Eq.(14), can be  studied numerically. An alternative is using the FPE which corresponds to this stochastic process.

\subsection{Fokker-Planck equation for the RDM}

For the RDM  
the probability density $P({\bf S},t)$ is defined as 
\begin{equation}
P({\bf S},t) = < \delta ({\bf S} - {\bf e}(t)) >_\epsilon
\end{equation}
where the angular brackets denote an average over all trajectories ${\bf e}(t)$ of Eq. (14).  

The probability density $P(\bf S, t)$ defined is a solution of the FPE obtained from Eq.(14). It reads 
\begin{eqnarray}
\frac{\partial {P(\bf S,t)}}{\partial t}   =   \nabla \cdot \Big[
  \frac {\mu_s}{\xi} {\bf S} \times \left({\bf S} \times { \bf {B} } \right)   \nonumber \\ - \, {\frac{1}{2\tau_B}} \, {\bf S} \times \left( {\bf S} \times \nabla \right)  \Big]P(\bf S,t). 
\label{e:FP_1}
\end{eqnarray}
For an elegant derivation of the FPE see for instance Ref.\cite{garanin}.
 The quantity $\tau_B$ introduced in Eq.(16) denotes the Brownian relaxation time,
\begin{equation}
\tau_B=\frac{3 \eta V_d}{{\kB}T} .
\end{equation}
Because the stochastic process conserves the length of $\bf e$ the probability density is of general form
\begin{equation}
P = \delta ( |{\bf S}| -1) \, Q({\bf S}, t).
\end{equation} 

The probability density $P(\bf S, t)$ contains all information about fluctuation averaged physical quantities. The averaged reduced magnetic moment, for instance, is given by
\begin{equation}
{\bf m} (t) = \int d^3 {\bf S} \, P({\bf S} , t) \, \bf S, 
\end{equation} 
the first moment of the probability density, which is equivalent to the average over thermal fluctuations, Eq.(7). 

It is easy to see that the Boltzmann distribution $P_0 \sim \mathrm{exp}({\mu_s\bf S} \cdot {\bf B}_0/k_BT)$ is an equilibrium solution of Eq.(16). In general, however, exact solutions of the FPE are not known  so that one has to rely on  numerical solutions of the  FPE or on approximations.

For a numerical solution of the FPE the construction of a fast and robust  algorithm  has been described previously. The important point is to discretize  the equation of motion for $Q$, Eq.(18),  in such a way that the normalization of $P$ is preserved independent of the mesh size.  For details the reader is referred to \cite{usa_FP}.

An approximate solution of the FPE, the effective field method,  has been described in \cite{mart,raik} the basic steps of which are outlined in the next section. 
   
\subsection{Effective field method}

The starting point is an ansatz for the probability density as in the works before, \cite{mart,raik}, assuming for $P({\bf S}, t)$  an expression of the form of an equilibrium density with adjustable  parameters, 
\begin{equation}
P = N^{-1} \delta ( |{\bf S}| -1) \, \mathrm {exp}( {\bf A}(t) \cdot {\bf S}) 
\end{equation} 
with normalization factor $N$ given by
\begin{equation}
N(t) = 4 \pi \frac{\mathrm{sinh}(A(t))}{A(t)}
\end{equation}
and $A(t) = \sqrt {( {\bf A}(t)\cdot {\bf A}(t))}$. 

The reduced magnetic moment resulting from this probability density, i.e. the first moment of the probability density,  is given by 

\begin{equation}
{\bf m}(t)  = L(A(t)) \frac {\fett{A(t)}} { { {A}}(t)}
 \end{equation}
where
\begin{equation} 
L(A) = \frac {1} { \mathrm {tanh} (A) } - \frac {1}{A}
\end{equation}
denotes the Langevin function. 
For the first moment of the probability density we obtain  from Eq.(16) after partial integration

\begin{eqnarray}
\int d^3 S {\bf S }\frac{\partial {P({\bf S},t)}}{\partial t} 
 =  \int d^3 S \Big[
  \frac {\mu_s}{\xi} {\bf S} \times \left({\bf S} \times { \bf {B} } \right)  \nonumber  \\ - \, {\frac{1}{2\tau_B}} \, {\bf S} \times \left( {\bf S} \times \nabla \right)  \Big]P(\bf S,t) .
\label{e:FP_1}
\end{eqnarray}
The  effective field $\bf A$ is determined by the requirement that this equation is fulfilled for the probability density $P(\bf S, t)$  defined in  Eq.(20)  leading to 
\begin{equation}
\frac {d \bf m}{dt} = -\frac {\mu_s}{\xi} \big [ \, (1-\frac {3 m}{A}) \, { ({\bf B} \cdot {\bf {\hat {m}}}) \, {\bf \hat m}} +( \frac {m}{A} -1) {\bf B} \big] - {\frac{1}{\tau_B}}\bf m.
\end{equation} 
$\bf {\hat m}$ denotes a unit vector in the direction of $\bf m$. Eqs. (22,23,25) are the basic equations of the EFM \cite{mart,raik}.  

Note that for a time independent  magnetic field $\bf B ={\bf B}_0 $ the equilibrium solution $\frac {d \bf m}{dt} = 0$ is parallel to $\bf{ B}_0$ so that we obtain from Eq. (25)  $A= \frac {\mu_sB_0}{\kB T}$. For the reduced magnetic moment we therefore obtain with Eq. (23)  
\begin{equation}
m = L(\frac {\mu _s B_0}{{\kB}T}),
\end{equation}
i.e. the equilibrium moment. 
This result is of course expected because the equilibrium probability density is of the same form as that assumed in Eq.(20). 

In the general case, however, because of the nonlinear relation between  $m(t)$ and $A(t)$  in Eq.(22), analytic solutions are not obvious. An exception is the special case of  a rotating magnetic field, Eq.(8), for which a stationary solution of Eq.(25) can be found. 
\subsection{Rotating magnetic fields}
For a rotating magnetic field  a stationary solution exists with time independent amplitude $m=\sqrt {(\bf m \cdot \bf m)}$ implying a time independent $A$, Eq.(22). 

To show this we note that for a time independent $m$ we have 
\begin{equation}
\fett m \cdot  \frac {d \bf m}{dt} = 0.  
\end{equation}
 Multiplying Eq.(25) with ${\bf m}$ we obtain
\begin{equation}
{\bf B}(t) \cdot {\bf {m}}(t) =\frac  {Am {\kB}T} {\mu_s}.
\end{equation}
Therefore a time independent $m$  (and $A$) requires a rotating magnetic moment $\bf m$  with constant phase lag, i.e.  $\bf m$  must be of the form 

\begin{equation}
 {\bf m}  = m \,({\mathrm {cos}}( \tilde \omega \,t -\phi) \,{\hat {\bf x}} + {\mathrm {sin}}( \tilde \omega \,t-\phi) \,{\hat {\bf y}} ).
 \label{e:t}
 \end{equation}
This form indeed solves  Eq.(25) and one finds after some algebra for the  phase lag
 \begin{equation}
 \phi ={\mathrm {arctan}}( \frac {2 \,\tilde  \omega \, \tau_B }{A/m-1})
\end{equation} 
and for the amplitude of the effective field
\begin{equation}
A = \frac{\mu_s B_0}{k_BT} \frac {(A/m-1) }{\sqrt  {  (2 \tilde \omega\tau_B)^2 + (A/m-1)^2}} 
\end{equation}
with
\begin{equation}
m = L(A).
\end{equation}
Eqs.(31-32) determine implicitly the quantities  $A$ and $m$.

We note in passing that the lag angel $\phi$ determines the energy absorption of the  MNPs by the rotating field. The absorbed power can be obtained from \cite{usov_0,usa_ROT,rosen}
 \begin{equation}
W = - \mu_s\langle {\bf e} \frac {d\bf B}{d t} \rangle
\end{equation}
so that we obtain from  Eqs.(8,29) 
\begin{equation}
W = \mu_s \tilde \omega B_0 \, m \, {\mathrm{sin}}(\phi).
\end{equation}

In the next section we will show that our analytic results, Eq.(30-32), are in nearly perfect agreement with those we obtained from numerical solutions of the FPE.    

 \begin{figure}[b] 
   \centering
 \includegraphics[width=2.9in]{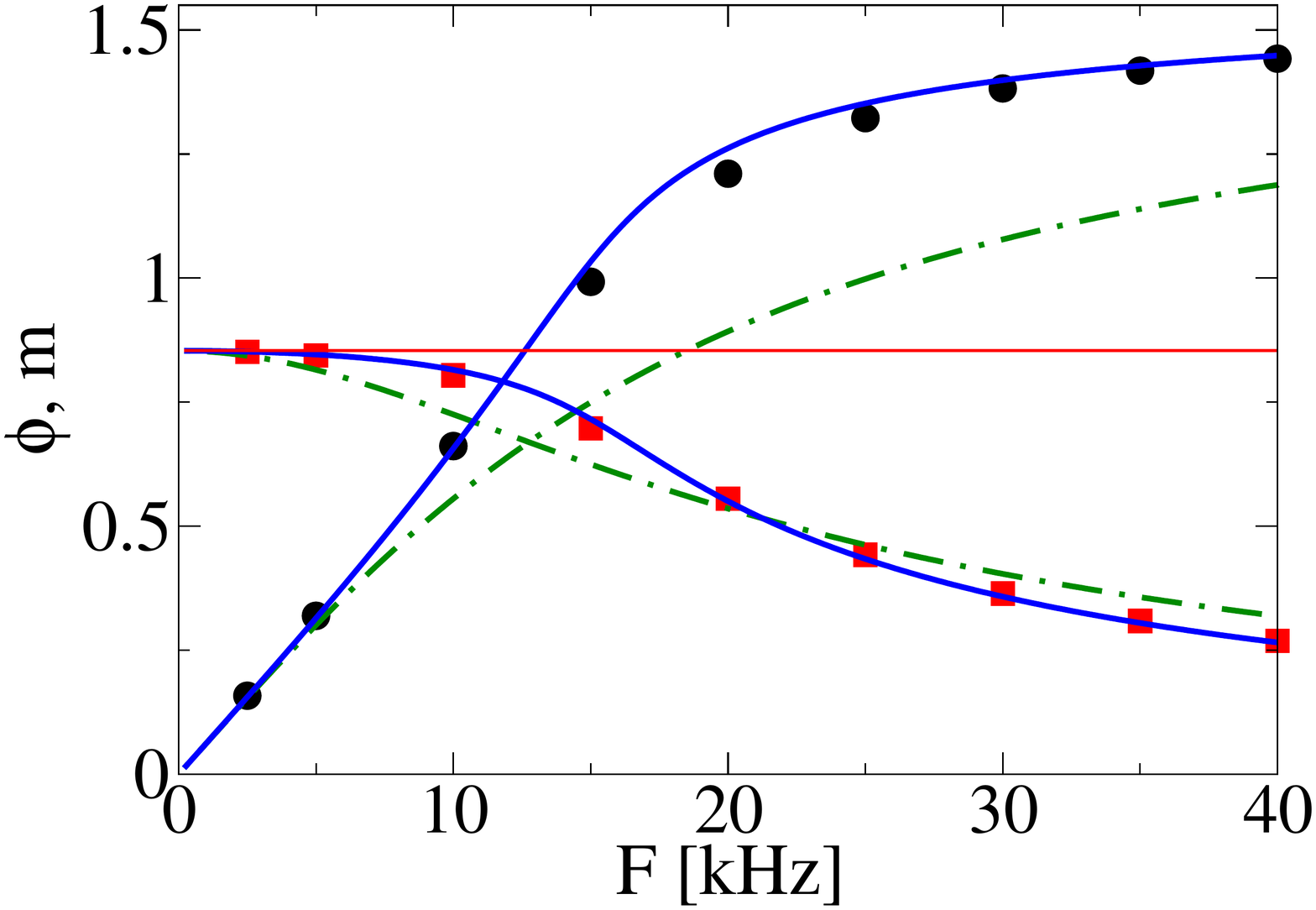}  \\[-0.3cm]
   \includegraphics[width=2.9in]{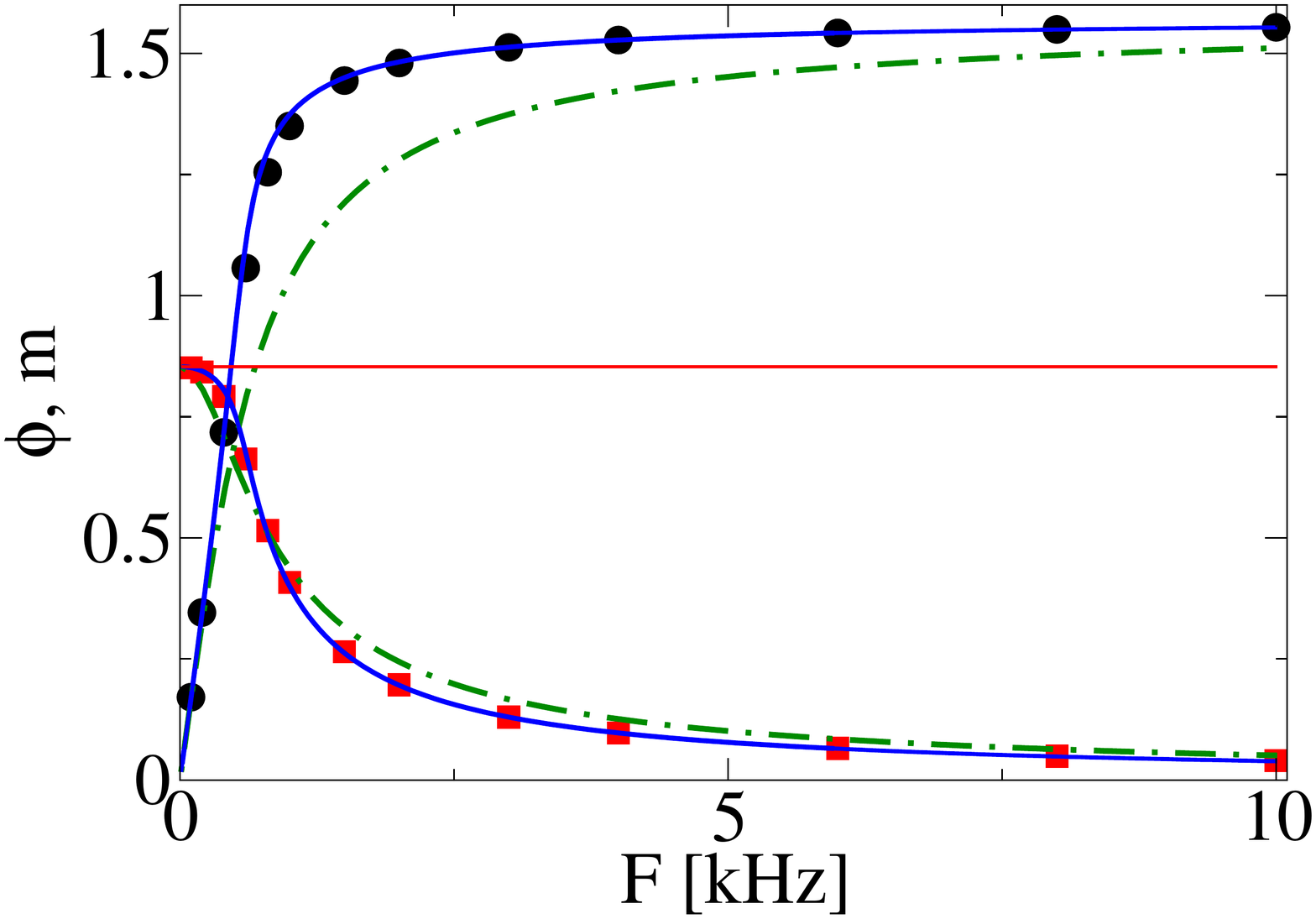}  \\[-0.3cm]
   \caption{(color online) 
   Phase lag $\phi$   and  magnetic moment $m$ versus frequency.\\
   Parameters: $\tilde\eta=1.0, T=300, r_{m}=15.0, B_0=5 $ mT. \\
   Upper panel: $r_d / r_m = 1.5$, lower panel $r_d /r_m =4.5$. \\
   Dots black and squares red: numerical solution of the FPE.
   Solid, blue: present paper,  dashed-dotted, green: Ref.\cite{ferro}.    }
   \label{f:fig_2}
   \end{figure}  

Different results  for the case that a rotating field is applied have been reported previously \cite{ferro}. Those results are based on expansions  of the EFM around the equilibrium solution \cite{mart,raik}. The following expressions for the phase lag and for the effective field -  called  $\xi$ for clarity - were obtained \cite{ferro}: 
 \begin{equation}
 {\phi}  = {\mathrm{ arctan}}( \tilde \omega \, \tau_B \frac {2 \, L(\xi)}{\xi-L(\xi)})
 \end{equation}
 with 
 \begin{equation}
 \xi =  \frac{\mu_s B_0}{k_BT}
 \end{equation}
 and 
 \begin{equation}
 m= L(\xi)\mathrm{cos}(\phi) .
 \end{equation}
 
The effective field $A$, Eq.(31),  is frequency dependent in general. For $\omega \rightarrow 0$ it  becomes identical to $\xi$, Eq.(36), so that in this limit the results obtained for the phase lag coincide. Differences are obvious for finite frequencies.

 Finally we would like to mention that for small fields the analytic results presented agree with each other and they also agree with results from linear response theory. The reason is simply that for small fields the quantity
 $Q(\bf S, t)$, Eq.(18), can be expanded as
 \begin{equation} 
 Q(\bf S, t) = 1+{\bf q}(t) \cdot \bf S + ...
 \end{equation}
where $\bf q$ is assumed to be linear in the field. Inserting this expression into Eq.(16) and keeping only terms linear in $B_0$ we arrive at a differential equation for ${\bf q}(t)$ which can be solved easily. It is important to note that this is an exact solution of the FPE (to linear order in $B_0$)  describing the stationary state. 

We do not present details of this calculation here because the results can also be obtained from an expansion of Eqs.(30-32) in linear order in $B_0$. The reason for this is that
for small  effective fields $P(\bf S,t)$, Eq.(20), agrees with Eq.(38) in linear order in $B_0$ so that the results for $\phi$ and $m$ obtained must coincide. 
 
\subsection{ Numerical analysis of the RDM}

 \begin{figure}[t] 
   \centering
    \includegraphics[width=2.9in]{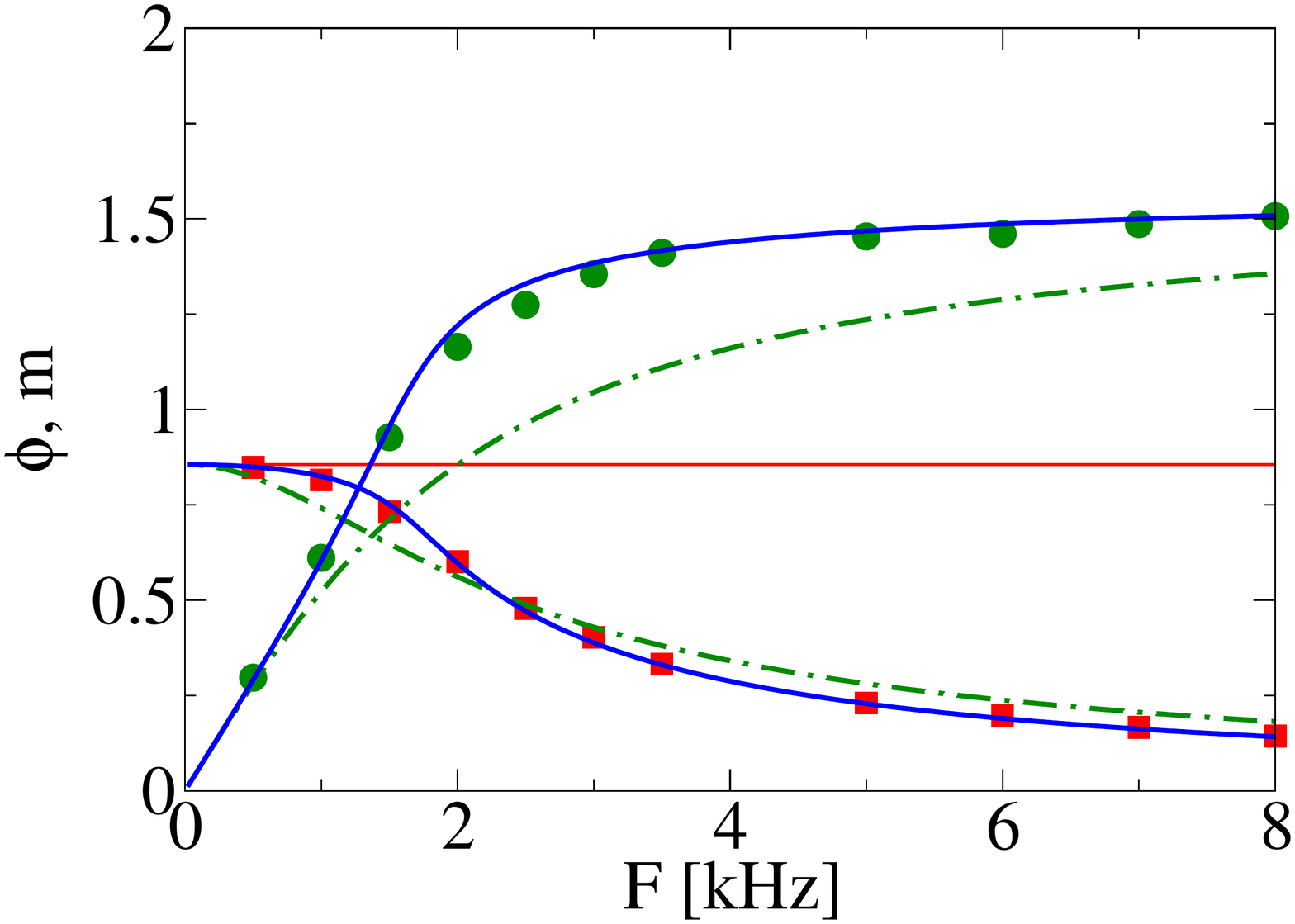}  \\[-0.3cm]
 \includegraphics[width=2.9in]{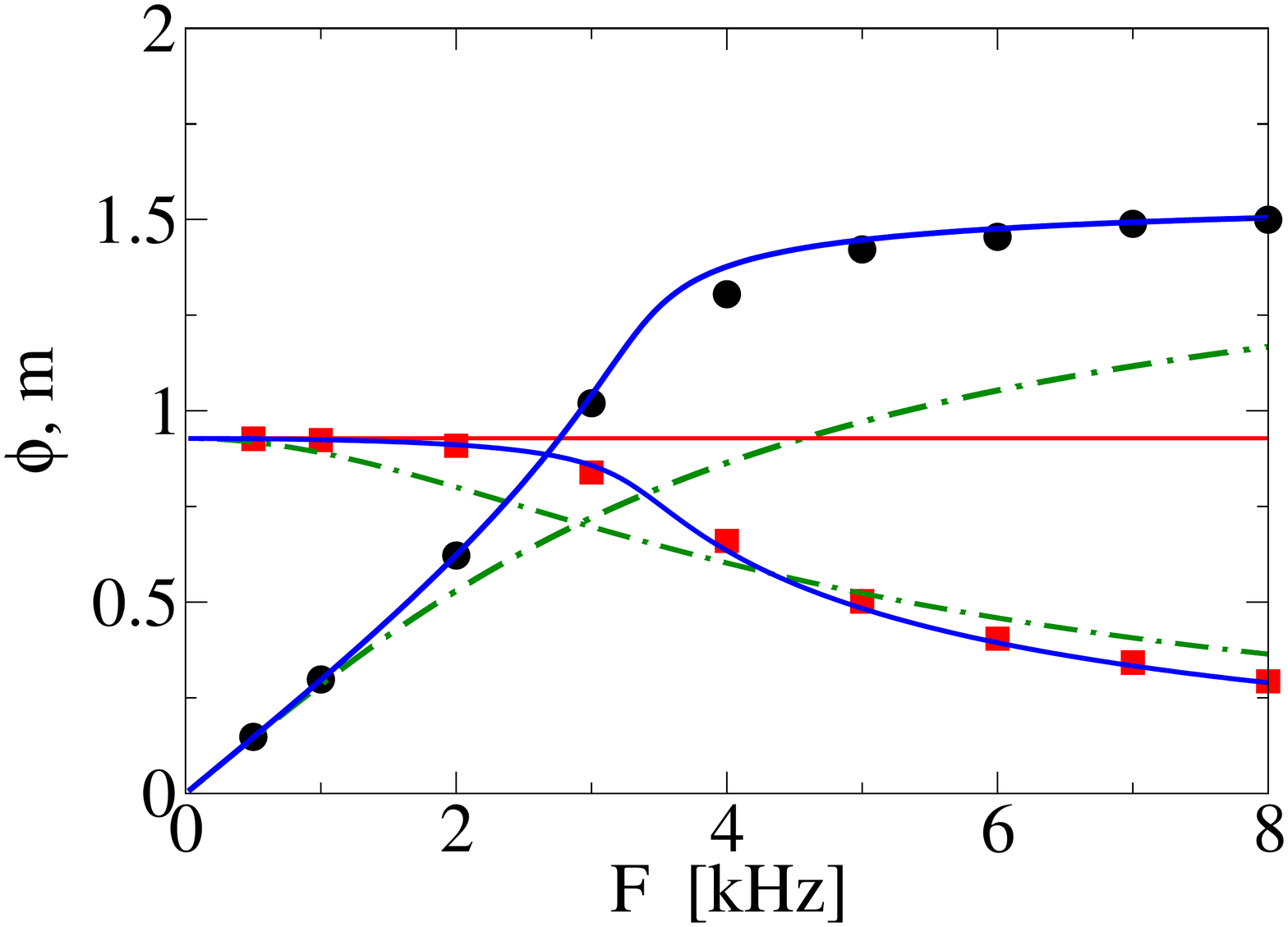}  \\[-0.3cm]
  \includegraphics[width=2.9in]{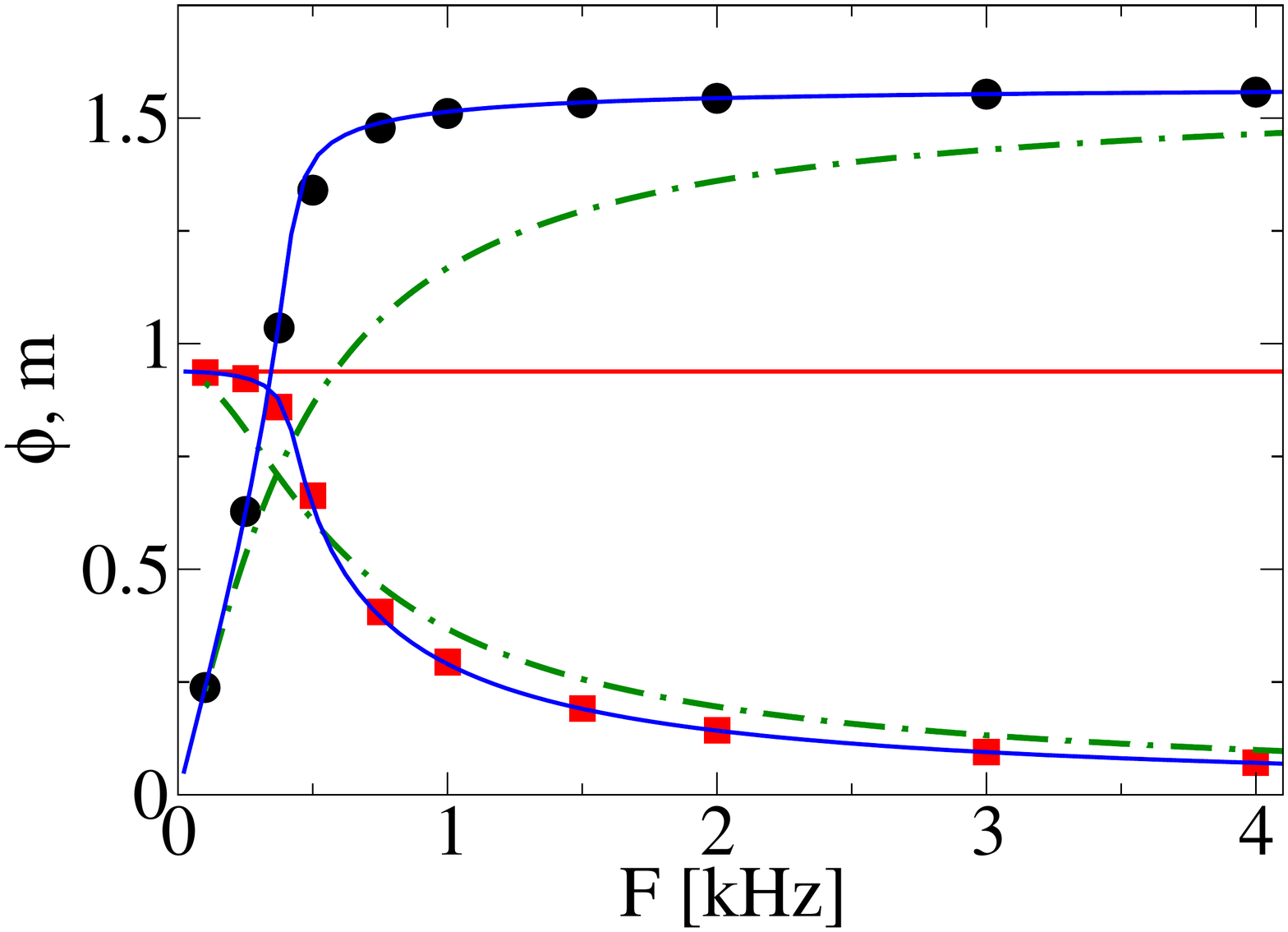}  \\[-0.3cm]
    \caption{(color online)  Phase lag $\phi$   and magnetic moment $m$ versus frequency. \\
    Parameters: $r_d / r_m = 2.5$, $r_{m}=19$ nm. \\
    Upper panel: $B=2.5$ mT, $\tilde \eta  = 1.0$,\\
    middle panel: $B=5$ mT. $\tilde \eta  = 1.0$,\\
    lower panel: $B=5$ mT. $\tilde \eta  = 8.0$.\\
     Dots black and squares red: numerical solution of the FPE.\\
  Solid blue: present paper,  dashed-dotted green: Ref.\cite{ferro}. }
      \label{f:fig_3}
   \end{figure}

Results obtained from a numerical solution of the FPE for the RDM will be compared in the following with those obtained from our analytic results, Eqs.(30-32) and also with those obtained in Ref. \cite{ferro} (Eqs. (35-37)). 

 From the numerical solution of the FPE in a  rotating field the magnetic moment ${\bf m }(t)$ is calculated using Eq.(19). After an initial time interval with length depending again strongly on the parameters of the system a stationary state with time independent phase lag is obtained.  Results for the phase lag and the reduced magnetic moment in the stationary state are discussed in the following.
    
   Fig.(2) shows the phase lag $\phi$  (black circles and ascending curves) and the magnitude of the induced magnetic moment, $m$, (red squares and descending curves) as function of frequency for $r_d/r_m =1.5$ (upper panel) and for $r_d/r_m=4.5$ (lower panel).  An increase of the dynamical radius $r_d$ increases the frictional torque which is compensated by the enormous increase of the lag angel observed at small frequencies (note the different scales in the abscissa of these two graphs).  This is consistent with the observation that an increase of the phase lag leads to an increase of the absorbed power according to Eq. (34).
 
 Phase lag and magnetic moment  calculated from the numerical  solution  of the FPE  are in very good agreement with our analytic results obtained from the  EFM  (solid lines, blue) in the entire frequency region shown. Results obtained from \cite{ferro} (dashed-dotted, green) deviate from the numerically exact result especially for larger frequencies.   
 
 The horizontal lines (solid, red) show the equilibrium magnetic moment for a field $B_0$, the amplitude of the rotating field. Obviously this result shows that for very small frequencies the system is in a state of quasi equilibrium  in which the magnitude of the magnetic moment is close to its equilibrium value rotating slowly.  With decreasing frequency the equilibrium value is approached asymptotically.

The results shown  are representative in the sense that  we always found an extremely good agreement between the results obtained from the FPE and those obtained from our analytic results based on the  EFM.

Fig.(3) shows results for particles with an increased magnetic radius, $r_m = 19$ nm. 
  An increase in the field strength from $B_0 = 2.5$ mT (upper panel) to $B_0 = 5$ mT (middle panel) leads to a reduction in the lag angel and to a significant increase in the magnetic moment in the low frequency region. This is explained by the increased driving torque exerted by the larger magnetic field. An increased frictional torque, on the other hand, leads to an increase of the lag angel and a decrease of $m$ as can be seen in the lower panel of Fig.(3) where the viscosity is enlarged by a factor of $8$ (note again the different scale in the abscissa of the figure in the lower panel). 

Finally, in Fig.(4)  we show results for the dependence of the lag angel and the magnetic moment on the ratio of the hydrodynamic to the magnetic radius, $r_d / r_m$, important for bioassay applications \cite{schritt2}.
Remarkable is the sharp increase of $\phi$ around  $r_d / r_m=2.5$. For larger values of $r_d / r_m$ the phase lag is nearly constant, i.e. insensitive to variations of $r_d$.

\begin{figure}[t] 
   \centering
 \includegraphics[width=3.0in]{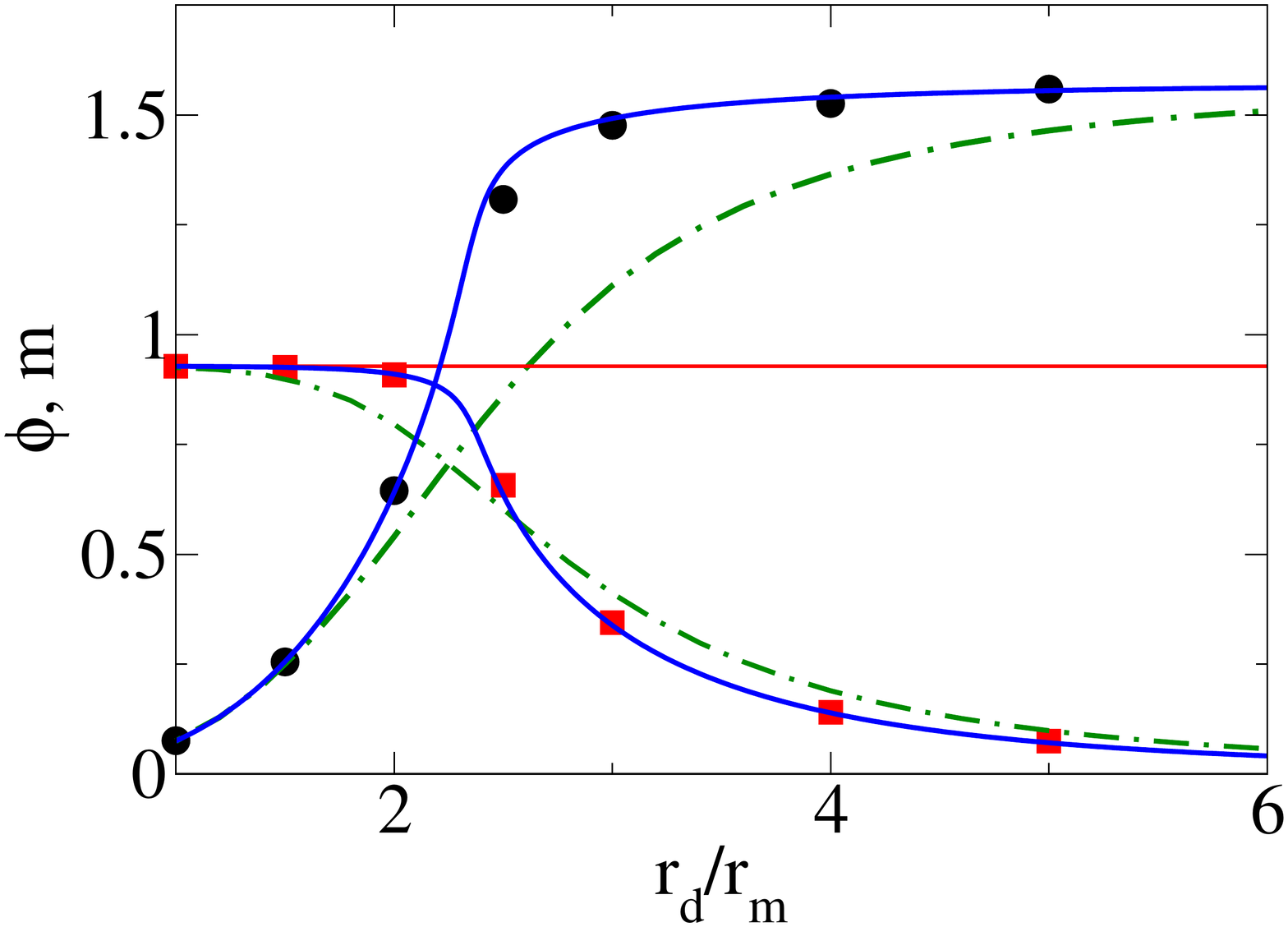}  \\[-0.3cm]
   \caption{(color online) 
  Phase lag $\phi$  and  magnetic moment $m$  versus  $ r_d / r_m $.\\
  Parameters: $\tilde\eta=1.0, T=300, r_m=19.0, B=5.0$  mT, $F = 4$ kHz.\\
    Dots black and squares red: numerical solution of the FPE.\\
  Solid blue: present paper,  dashed-dotted green: Ref.\cite{ferro}.}       
   \label{f:fig_4}
   \end{figure}

   \section{Conclusions}

Our numerical calculations based on the kinetic equations for MNPs dissolved in a viscous liquid show that under the influence of a rotating magnetic field a transition takes place from a state with magnetic moment locked to the anisotropy axis of the MNP to a state with free rotation of the moment depending on the anisotropy energy. This scenario is expected physically and our results support this picture quantitatively. From these investigations we  can conclude that  for MNPs  at room temperature with  magnetic parameters typical for  iron oxides (magnetite) the moment can be considered as being locked if  the magnetic radius $r_m$ is larger than about $12$ nm, c.f. Fig.(1). Particles with a larger radius can be described within the RDM.

The EFM which is based on the FPE for the RDM has been reconsidered and it has been applied to the dynamics of MNP in rotating magnetic fields. The analytic results we obtain   
are shown  to be in perfect agreement with numerical solutions of the FPE in the range of parameters studied. Therefore these results can be used with success in applications avoiding lengthy calculations on the basis of the FPE.

\end{document}